\pdfoutput=1
\documentclass{article}
\usepackage{spconf,amsmath,graphicx}
\usepackage{multirow}
\usepackage{siunitx}

\title{zero-shot text-to-speech synthesis \\conditioned using self-supervised speech representation model}
\name{Kenichi Fujita, Takanori Ashihara, Hiroki Kanagawa, Takafumi Moriya, Yusuke Ijima}
\address{NTT Corporation, Japan}
\begin{document}
\ninept
\maketitle
\begin{abstract}
This paper proposes a zero-shot text-to-speech (TTS) conditioned by a self-supervised speech-representation model acquired through self-supervised learning (SSL). Conventional methods with embedding vectors from x-vector or global style tokens still have a gap in reproducing the speaker characteristics of unseen speakers. A novel point of the proposed method is the direct use of the SSL model to obtain embedding vectors from speech representations trained with a large amount of data. We also introduce the separate conditioning of acoustic features and a phoneme duration predictor to obtain the disentangled embeddings between rhythm-based speaker characteristics and acoustic-feature-based ones. The disentangled embeddings will enable us to achieve better reproduction performance for unseen speakers and rhythm transfer conditioned by different speeches. Objective and subjective evaluations showed that the proposed method can synthesize speech with improved similarity and achieve speech-rhythm transfer. 

\end{abstract}
\vspace{-3.0mm}
\section{Introduction}
\label{sec:intro}
\vspace{-2.5mm}
Developments in text-to-speech (TTS) synthesis have achieved natural-sounding human speech from a large amount of speech uttered by a single speaker~\cite{shen2018natural,ren2020fastspeech}. Through multi-speaker TTS, arbitrary speaker's TTS has been achieved from a small amount of target speaker utterances~\cite{cooper2020zero}. However, re-training the acoustic models with adaptation data, which requires computational resources and time consuming training, is still necessary for obtaining high-quality synthesized speech by an arbitrary speaker. This implies that we encounter difficulty with zero-shot TTS, adapting an acoustic model by using only a limited amount of data without re-training. This paper addresses zero-shot TTS in a severe situation.

Most zero-shot TTS methods are based on neural speaker embeddings, i.e., the continuous vector representations of speaker information. Two main approaches have been used to obtain such embeddings: speaker-recognition-based embedding such as d-vector~\cite{heigold2016end,doddipatla2017speaker} and x-vector~\cite{cooper2020zero, snyder2018x}, and embedding from the reference encoder trained simultaneously with the acoustic model, such as using global style tokens (GSTs)~\cite{wang2018style, valle2020mellotron}. The advantage of the first approach is obtaining robust embeddings for unseen speakers because a large amount of data not limited to the training data for TTS can be used for training the embedding extractor. However, individual speech rhythm, an important factor among speaker characteristics~\cite{zetterholm2002intonation,zetterholm2003same,gomathi2012analysis} can not be sufficiently reflected on the synthesized speech using the embeddings. This is because the embeddings mainly carry acoustic-feature-based characteristics not rhythm-based ones~\cite{fujita21_interspeech}. The second method, i.e., using GSTs, achieves the reference speech's style modeling to some extent. The problem is the limited reproduction quality for unseen speakers because GSTs must be trained only with a comparably small amount of data for TTS. The goal of this study was zero-shot TTS, which can synthesize a more natural reproduction for the unseen target speaker's speech.

To achieve this goal, we believe that self-supervised learning~(SSL) models such as HuBERT~\cite{hsu2021hubert} and wav2vec 2.0 (w2v2) ~\cite{baevski2020wav2vec}, which obtain speech representations from a large amount of data without any labels, are promising. SSL models have been widely used in many speech-research areas, such as automatic speech recognition (ASR), speaker recognition, and speech-emotion recognition~\cite{yang21c_interspeech}. For speech-generation tasks, SSL models have been used as the robust phoneme-like unit extractor for high-quality speech re-synthesis~\cite{polyak21_interspeech}, voice conversion (VC)~\cite{lin2021fragmentvc}, and direct speech-to-speech translation~\cite{lee-etal-2022-direct}. Other studies have shown interesting properties of SSL models, e.g., the extracted representation vectors of each layer have different information related to the input speech sequence~\cite{chen22g_interspeech,shor2022universal}. Therefore, we expect embeddings including richer information well-suited for zero-shot TTS can be obtained from an SSL model because such a model is trained with a large amount of data containing many speakers and speaking styles. With these embeddings, the reproduction and quality of zero-shot TTS should greatly improve.

We propose a zero-shot TTS method conditioned using an SSL model. The key idea is the direct use of an SSL model to obtain speaker embeddings in the same manner as models in \cite{yang21c_interspeech}. By obtaining the embedding vector expressed as the weighted-sum of outputs from each hidden layer of the pre-trained SSL model, the proposed method obtains an adequate embedding vector from a massive amount of SSL parameters for reproducing speaker characteristics in zero-shot TTS. We also introduced the separate conditioning of acoustic features and a phoneme duration predictor by obtaining embedding vectors for the duration predictor and other predictors in the non-autoregressive TTS model, respectively. This separate conditioning improves the zero-shot TTS performance by obtaining the disentangled embeddings between rhythm-based speaker characteristics and acoustic-feature-based ones. The disentangled embeddings also enable speech-rhythm transfer, which generates speech with one target speaker's acoustic characteristics and another target speaker's rhythm characteristics. Audio samples are available in our demo page\footnote{https://ntt-hilab-gensp.github.io/23icassp-sasb-zeroshot/}.


\begin{figure}[tb]
    \centering
    \includegraphics[width=0.7\linewidth]{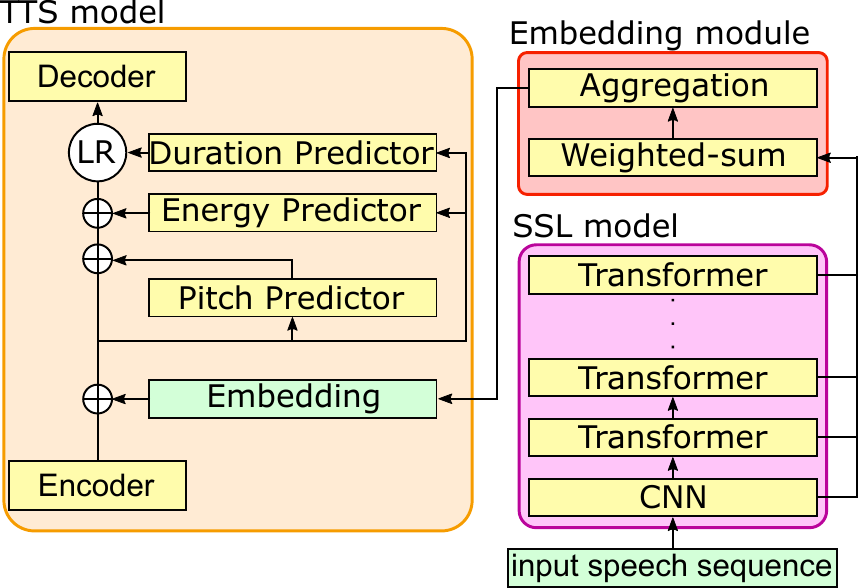}
    \vspace{-4mm}
    \caption{Overview of the proposed method. Non-autoregressive TTS model is conditioned with latent representations from SSL model. LR denotes length regulator.}
    \label{fig:overview}
\end{figure}
\vspace{-3.5mm}
\section{Proposed method}
\label{sec:model_architecture}
\vspace{-3.5mm}
\subsection{Overview}
\vspace{-2.5mm}
Figure~\ref{fig:overview} shows an overview of the proposed method. The method mainly consists of three components, i.e., a non-autoregressive TTS model, such as FastSpeech2~\cite{ren2020fastspeech}, SSL model, and embedding module. Although these components are similar to other zero-shot TTS methods, the main difference is the use of an SSL model to obtain the embedding vector instead of other embedding extractors, such as x-vector. Although GenerSpeech~\cite{huang2022generspeech} uses an SSL model as the global style estimator, the proposed method is fundamentally different because GenerSpeech can be considered as an extension of speaker-recognition-based neural speaker embeddings. This requires fine-tuning the SSL model for speaker and emotion recognition.

The proposed method first converts the input speech sequence into a frame-level sequence of speech representation vectors: the outputs from each layer of the SSL model on every frame. The dimension of the speech representation vector is $D_{sr} = D_l \times L$, and the frame-level sequence of the speech-representation vector consists of $D_{sr} \times F$. Here, $D_l$, $L$, and $F$ are the dimension of each layer, number of layers of the SSL model, and number of frame-level sequences, respectively. The frame-level sequence of the speech-representation vector is also converted into a fixed length vector, i.e., embedding vector, through the embedding module described in Sect.~\ref{sec:SSL_embedding_ext}. Finally, the obtained embedding vector is input into the non-autoregressive TTS model. Since the non-autoregressive TTS model and embedding module are trained simultaneously during model training, a suitable embedding vector can be obtained from a massive amount of SSL parameters for the TTS model. Details of the proposed method are described in the following sections.

\vspace{-3.5mm}
\subsection{Embedding module}\label{sec:SSL_embedding_ext}
\vspace{-2.5mm}
The embedding module converts the output frame-level sequence of the speech-representation vector into a fixed-length embedding vector. This module consists of two parts, i.e., weighted-sum and aggregation. The weighted-sum part converts the sequence of the speech-representation vector into the sequence of the weighted-sum speech representation vector, which consists of $D_l \times F$, in the same manner as models in \cite{chen22g_interspeech}. The aggregation part then converts the obtained frame-level sequence into a fixed-length embedding vector.

As the aggregation part, we compare two methods, i.e., average pooling and soft-attention of LSTM outputs. Average pooling obtains the average vector of the frame-level sequence in the same manner as in previous studies. Although it is a simple approach to obtain an embedding vector, the simple averaging process ignores temporal features, e.g., speech rhythm. 

Soft-attention of LSTM outputs, on the other hand, is mainly composed of two steps: 1)~LSTM processes the weighted-sum speech-representation vector, and 2)~the sequence of the LSTM hidden state is aggregated with an attention-based structure~\cite{bhattacharya2017deep, ando2018soft}. The attention mechanism obtains embedding vectors more suitable for reproducing the speaker characteristics with the non-autoregressive TTS model by extracting important frames from speech-representation sequences. This method will also enable to capture temporal features, e.g., speech rhythm, by taking the time series of the weighted-sum speech representation vector into account. This is because the speaking rhythm information may be considered as the velocity and acceleration of representation-vector sequences from SSL models.


\begin{figure}[tb]
    \centering
    \includegraphics[width=0.7\linewidth]{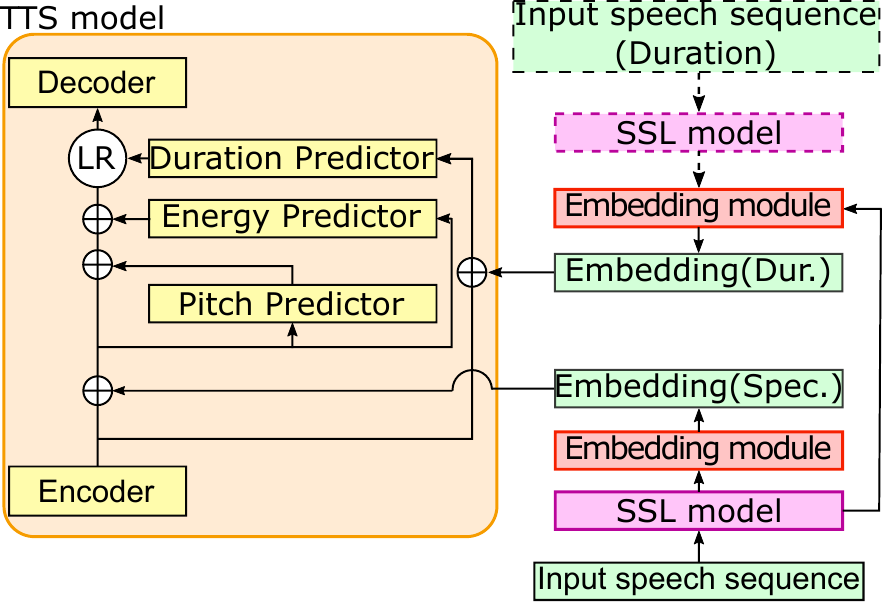}
    \vspace{-4mm}
    \caption{Overview of separate conditioning. Duration predictor and other components are conditioned separately. For rhythm transfer, different input speech sequence is given for conditioning duration predictor (indicated with dotted boxes).}
    \label{fig:overview_sep}
\end{figure}
\vspace{-3.5mm}
\subsection{Separate conditioning}\label{sec:independet_conditioning}
\vspace{-2.5mm}
Figure~\ref{fig:overview_sep} gives an overview of the separate conditioning of acoustic features and a phoneme duration predictor. Because the output of the duration predictor does not affect other predictors, a phoneme duration predictor and other predictors would be independent in most non-autoregressive TTS models. Therefore, the proposed method can condition both the duration predictor and other predictors by obtaining embedding vectors for each predictor separately. This leads to precise speaker modeling by obtaining disentangled embedding vectors containing rhythm-based speaker characteristics and acoustic-feature-based ones separately for each predictor.

These disentangled embedding vectors also enable speech-rhythm transfer, which can generate speech with the acoustic characteristics of a target speaker and rhythm characteristics of another target speaker. The non-autoregressive TTS model generates speech parameters referring to different speeches given at each predictor.

\vspace{-3mm}
\section{Experimental Setup}
\vspace{-2.5mm}
\subsection{Training-set for TTS model}
For training the non-autoregressive TTS model, we used an in-house Japanese speech database which includes 1,083 speakers. This database consists of several speaker types including professional speakers, i.e., newscasters, narrators, and voice actors, and non-professional speakers. The database was split into three; 135,202 utterances by 978 speakers, 7,243 by 52, and 6,421 by 53 for training, validation, and test. The sampling frequency of the speech was \SI{22.05}{\kHz}. All speech samples were manually annotated with the accentual information. 
\vspace{-3mm}
\subsection{Training conditions}
\vspace{-2.5mm}
We newly trained two Japanese SSL models, i.e., w2v2 and HuBERT based on the basis of fairseq~\cite{ott2019fairseq}. The training data were from the corpus of spontaneous Japanese (CSJ)~\cite{maekawa2003corpus}, which includes 660 hours of speech data uttered by 1,417 Japanese speakers. The model parameters and training procedures were the same as {\sc BASE} in the original w2v2 and HuBERT studies~\cite{hsu2021hubert, baevski2020wav2vec}. An SSL model processes the input \SI{16}{kHz} raw audio sequence into 768-dimensional sequences, and the embedding module converts them into a 256-dimensional fixed-length vector. The parameters of these SSL models were fixed while training the non-autoregressive TTS model, i.e., FastSpeech2 implemented on the basis of a previous study~\cite{chien2021investigating}. We use the Adam optimizer and follow the same learning
rate schedule in Vaswani et al.~\cite{NIPS2017_3f5ee243}. The input and target sequence to FastSpeech2 were a 303-dimensional linguistic vector and 80-dimensional melspectrograms. The frame shift was \SI{5.0}{\ms}. To evaluate the performance by changing each condition, we trained FastSpeech2 with different SSL models, i.e., w2v2 and HuBERT, conditionings (common and separate), and aggregation methods (average pooling, and LSTM with attention). ``Common'' and ``separate'' conditionings respectively indicate a common conditioning for each predictor and separate conditioning described in Sect.~\ref{sec:independet_conditioning}.

We also trained FastSpeech2 with x-vector trained using CSJ as the conventional method (hereafter, x-vector) to compare with the proposed method. The 256-dimensional x-vector was extracted from \SI{16}{kHz} speech signals by using the speaker identification model based on FastResNet-34 with self-attentive pooling (SAP) trained with angular prototypical loss~\cite{chung2020in}. As well as the proposed method, we trained x-vector models with two conditions: common and separate. Under the ``separate'' condition, x-vectors were converted with a full-connected layer for each predictor. We used HiFi-GAN~\cite{NEURIPS2020_c5d73680} for waveform generation for all proposed and x-vector methods.


\setlength{\tabcolsep}{1.5mm} 
\begin{table}[tb]
\caption{Results of objective evaluations. Common and separate are conditions with single and separate embeddings, respectively. Spec. and Dur. are MAE of log melspectrogram and RMSE of phoneme duration (ms). }
\label{tab:obeval_cSJ}
\begin{tabular}{|c|c|c||cc||cc|}
\hline
\multirow{2}{*}{model}  & \multirow{2}{*}{aggregation} & \multirow{2}{*}{condition} & \multicolumn{2}{c||}{parallel}      & \multicolumn{2}{c|}{non-parallel}  \\ \cline{4-7} 
                          &          &                          & \multicolumn{1}{c}{Spec.}  & Dur.   & \multicolumn{1}{c}{Spec.}  & Dur.   \\ \hline\hline
\multirow{2}{*}{x-vector} & \multirow{2}{*}{SAP}& common        & \multicolumn{1}{c}{1.33} & 21.4 & \multicolumn{1}{c}{1.36} & 22.0 \\ 
                          &          &          separate        & \multicolumn{1}{c}{1.32} & 21.8 & \multicolumn{1}{c}{1.34} & 22.5 \\ \hline 
\multirow{4}{*}{w2v2}     & \multirow{2}{*}{average}  &  common & \multicolumn{1}{c}{1.19} & 17.8 & \multicolumn{1}{c}{1.33} & 22.5 \\ 
                          &  &                          separate& \multicolumn{1}{c}{1.20} & 18.2 & \multicolumn{1}{c}{1.33} & 22.8 \\ \cline{2-7} 
                          & \multirow{2}{*}{LSTM} & common      & \multicolumn{1}{c}{1.19} & 18.2 & \multicolumn{1}{c}{1.30} & {\underline{\textbf{21.3}}} \\ 
                          &  & separate                         & \multicolumn{1}{c}{1.19} & 16.0 & \multicolumn{1}{c}{1.30} & 22.9 \\ \hline 
\multirow{4}{*}{HuBERT} & \multirow{2}{*}{average}   & common   & \multicolumn{1}{c}{1.17} & 17.3 & \multicolumn{1}{c}{1.30} & 22.3 \\ 
                          & &   separate                        & \multicolumn{1}{c}{1.16} & 17.0 & \multicolumn{1}{c}{1.30} & 22.4 \\ \cline{2-7} 
                          & \multirow{2}{*}{LSTM}&common        & \multicolumn{1}{c}{\underline{\textbf{1.15}}} & 17.7 & \multicolumn{1}{c}{\underline{\textbf{1.27}}} & 21.6 \\ 
                          & &    separate                       & \multicolumn{1}{c}{\underline{\textbf{1.15}}} & \underline{\textbf{15.6}} & \multicolumn{1}{c}{\underline{\textbf{1.27}}} & 22.0 \\ \hline
\end{tabular}
\end{table}

\vspace{-2mm}
\section{Results}
\vspace{-2mm}
\subsection{Objective evaluation on zero-shot TTS}\label{sec:objective_eval}
We first conducted an objective evaluation to evaluate the performance of the proposed and conventional (x-vector) methods under a data-parallel condition, in which the text to synthesize matches the text of the reference speech, and data-non-parallel condition in which the text to synthesize does not match the reference. Under the non-parallel condition, we randomly selected one utterance from each speaker as a reference speech. The mean absolute error (MAE) of the log melspectrogram and root mean square error (RMSE) of phoneme duration were used as the evaluation metrics. To calculate the MAE between generated and test ones with the same time alignment, we generated a log melspectrogram using the original phoneme durations extracted from test speech. We also obtained the RMSEs of durations by comparing the original phoneme durations of test speech with predicted ones from the duration predictor. Note that under the non-parallel condition, objective metrics were obtained not by comparing the target speaker's reference speech and the generated one but by comparing the target speaker's speech and the generated one with the same speech content.

Table~\ref{tab:obeval_cSJ} lists objective-evaluation results. Under the parallel condition, the proposed method performed better for both melspectrogram and phoneme duration than the x-vector. Comparing the embedding and aggregating methods, the separate conditioning with LSTM-based aggregation performed the best with both w2v2 and HuBERT. Since the RMSEs of phoneme duration drastically improved using the separate conditioning with LSTM-based aggregation, the proposed method can successfully acquire embeddings having rhythm-based speaker characteristics and acoustic-feature-based ones separately.

Under the non-parallel condition, the proposed method performed almost comparably to x-vector in phoneme duration, although the MAEs of the melspectrogram improved. The reason for this would be the intra-speaker variation. Even though the same speaker utters, the speech rhythm of each utterance would be inconsistent. Therefore, the proposed method would not necessarily lead to performance improvement in the non-parallel case, although it can accurately reproduce the characteristics of the reference speech.

HuBERT performed better than w2v2, which is consistent with the results in a previous study~\cite{yang21c_interspeech}, where HuBERT performed well in many tasks.

\begin{table}[tb]
  \caption{Naturalness and similarity scores with 95\% confidential interval.}
 \vspace*{-2mm}  
  \label{table:MOS}
  \begin{center}
  \begin{tabular}{ccc}
  \hline
  \noalign{\vskip.5mm}
    Model& MOS-naturalness& DMOS-similarity\\ 
  \hline
  x-vector &$3.40\pm0.05$&$2.86\pm0.05$\\
  HuBERT (average) &$3.25\pm0.05$&$\mathbf{3.73\pm0.04}$\\
  HuBERT (LSTM) & $\mathbf{3.45\pm0.05}$&$\mathbf{3.73\pm0.04}$\\
  \noalign{\vskip.5mm}
  \hline
  \end{tabular}%
  \end{center}
 \vspace*{-3.5mm}    
\end{table}

\vspace{-3mm}
\subsection{Subjective evaluations on zero-shot TTS}~\label{sec:subjective_eval}
We conducted subject evaluations to evaluate the naturalness and similarity of the proposed method. From the results of objective evaluations, we compared the conventional method, i.e.,  separately conditioned x-vector, with the separately conditioned proposed method, i.e., using HuBERT with average and LSTM aggregation (HuBERT~(average) and HuBERT~(LSTM)). For each model, we synthesized 20 sentences from each of the six speakers in test data (Male and Female children, adults, and voice actors) under the non-parallel condition. To evaluate the performance regarding out-of-domain (OOD) speakers, children speakers were chosen as such speakers for the SSL models and x-vector. This is because CSJ mainly includes younger adults. Fifteen participants rated the naturalness of synthetic speech on the basis of the mean opinion score (MOS) on a five-point scale of 5: very natural to 1: very unnatural. The similarity was rated on the basis of the differential MOS (DMOS) on a five-point scale of 5: very similar to 1: very dissimilar.

Table~\ref{table:MOS} shows that x-vector and HuBERT (LSTM) are almost the same naturalness, better than HuBERT (average), and both HuBERT versions have higher similarity to the target speech than x-vector. There are two reasons for the better performance of the proposed method: the performance of rhythm reproduction and robustness for OOD speakers. As shown in objective evaluations, the proposed method can capture rhythm-based speaker characteristics, which is one crucial factor for perceived speaker similarity. Therefore, the proposed method makes it possible to generate speech with higher similarity than x-vector, especially for speakers with characteristic speech rhythms, i.e., voice actors. The similarity scores of x-vector for OOD speakers, i.e., children, were also lower than those with the proposed method. One reason for this is that x-vector cannot adequately distinguish OOD speakers. Since the x-vectors obtained from different children speakers would be close in the embedding space, x-vector would not be able to reproduce speaker characteristics for children adequately. In contrast, the proposed method captured speaker characteristics for OOD speakers by obtaining embedding vectors from a large number of SSL parameters.

\begin{table}[tb]
  \caption{Preference scores for similarity on speech-rhythm transfer.}
  \label{table:pref_rhythm}
  \begin{center}
  \begin{tabular}{cccc}
  \hline
  \noalign{\vskip.5mm} 
    Model& preference
    \\ 
  \hline
  x-vector vs HuBERT &9.3\% - 90.7\%\\
  \noalign{\vskip.5mm}
  \hline
  \end{tabular}%
  \end{center}
\end{table}

\begin{table}[tb]
 \vspace*{-6.0mm}   
  \caption{Speaking rate of original, reference, and generated utterances (mora/sec) with standard deviation. Original and reference (Dur.) are speaking rate of speaker and speaker of reference speech given for duration predictor, respectively. }
  \vspace*{-4.5mm}
  \label{table:mora_sec}
  \begin{center}
  \begin{tabular}{c|cc|cc}
  \hline
  \noalign{\vskip.5mm} 
    Speaker & original &reference (Dur.) &x-vector &HuBERT
    \\ 
  \hline
  \#1 & 8.17 & 5.75 & $7.68\pm0.04$ & $5.64\pm0.05$\\
  \#2 &5.75 & 8.17 & $7.81\pm0.05$ & $8.13\pm0.05$\\ 
  \noalign{\vskip.5mm}
  \hline
  \end{tabular}%
  \end{center}
 \vspace*{-4.5mm}    
\end{table}
\vspace{-3mm}
\subsection{Evaluations on speech-rhythm transfer}
\vspace{-2.5mm}
We conducted an XAB test to evaluate the performance of the speech-rhythm transfer under the non-parallel condition with the proposed and conventional methods (separately conditioned HuBERT (LSTM) and x-vector). We selected two speeches from two female speakers \#1 and \#2 from the test data, respectively. Twenty utterances of them were generated by replacing their speaking rhythm with each other. In case an input-text X is synthesized with the voice of speaker \#1, the reference speech and reference speech (duration) are respectively from speaker \#1 and \#2. Text X and speech contents are all different. The participants were the same as in the previous subjective evaluations. Each was presented with synthesized speech samples then asked which sample had a similar rhythm to the reference speech. As the reference speech, we used one utterance used for conditioning the duration predictor. All permutations of synthetic speech pairs were presented in two orders (XAB and XBA) to eliminate bias in the order of stimuli.


Tables~\ref{table:pref_rhythm} and \ref{table:mora_sec} respectively list the preference scores and speaking rates of the original and generated utterances. The speaking rates of the original and reference (Dur.) are  those of their reference speech, and those of the generated are the average of the 20 generated utterances from each method. The proposed method had a higher preference score and closer speaking rate to the reference speech than x-vector. These results indicate that the proposed method enables rhythm transfer. However, x-vector could not reflect speech rhythm because it mainly carries acoustic-feature-based speaker characteristics.


\vspace{-5.5mm}
\section{Discussion}
\vspace{-4.0mm}
\subsection{Contribution analysis for weighted-sum}
\begin{figure}[tb]
    \centering
    \includegraphics[width=1.0\linewidth]{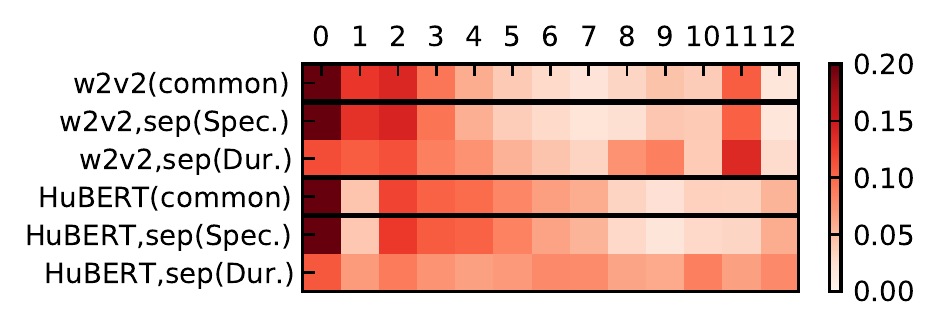}
    \vspace{-9.9mm}
    \caption{Visualized weight that aggregates representations from SSL models. ``sep'' is separate condition. ``spec'' and ``dur'' are weights for spectrogram and duration, respectively.}
    \label{fig:heatmap}
\end{figure}

\vspace{-3.mm}
To analyze the contribution of each layer in the SSL models, we visualize the weights that aggregate representations (Fig.~\ref{fig:heatmap}). We visualized the weights of LSTM aggregation methods of w2v2 and HuBERT with common and separate conditioning. The first and second rows of separate model show the weights for conditioning the acoustic features and phoneme duration, respectively. 

We can see that the zeroth layer, the output from the convolutional neural network, is dominant when the model is conditioned with the common condition. This indicates that the model extracts speaker information from the layers, including the information close to the spectrogram~\cite{chen22g_interspeech,shor2022universal}. On the other hand, the weights of the separate condition show interesting tendencies. The weights for conditioning acoustic features (Spec.) show almost the same tendency as common condition ones. However, the weights for the phoneme duration (Dur.) weigh on different layers comparably deeper layers. These tendencies indicate that the dominant factor of obtained embeddings under the common condition would still be acoustic feature-based speaker characteristics as well as x-vector. In other words, the proposed method with separate conditioning can successfully extract disentangled embeddings between rhythm-based speaker characteristics and acoustic-feature-based ones by obtaining different information from the SSL model suited for conditioning acoustic features and the phoneme duration predictor, respectively.


\vspace{-3.5mm}
\subsection{Analysis of training language and model size}
\vspace{-1.mm}
We finally analyzed the language dependency and effects of SSL model size. We used four publicly available SSL models\footnote{https://github.com/facebookresearch/fairseq}, w2v2 and HuBERT {\sc BASE} trained on LibriSpeech, and w2v2 and HuBERT {\sc LARGE} trained on LibriLight~\cite{hsu2021hubert,baevski2020wav2vec}. LibriSpeech and LibriLight are English speech data corpora of 960 hours from 2,338 speakers and 60K hours from 7,439 speakers, respectively. 


Table~\ref{table:obj_multi_large} lists the objective evaluation results in the same manner as in Table~\ref{tab:obeval_cSJ}. We first analyzed the language dependency by comparing SSL models trained with CSJ and LibriSpeech both having almost the same amount of training data. The results indicate that the language dependency on the proposed method would be limited since the performance of both models was comparable. As shown in the previous section, acoustic-feature-based speaker characteristics are obtained from the shallower layer, which is generally not strongly related to language. Additionally, the speaking rhythm information would not be related to language because this information would be extracted from the velocity and acceleration of representation-vector sequences from the SSL models.

Next, we analyzed the effect by the amount of training data and model parameters. We can see that the {\sc LARGE} model trained with larger data achieved higher reproduction in terms of melspectrogram. By using a larger variety of speakers, the SSL model would be able to extract the characteristics of unseen speakers more robustly.

\begin{table}[tb]
\vspace{-3mm}
\caption{Results of objective evaluations with SSL models trained with English corpus.}
\label{table:obj_multi_large}
\begin{tabular}{|c|c||cc||cc|}
\hline
\multirow{2}{*}{model}  & \multirow{2}{*}{corpus}  & \multicolumn{2}{c||}{parallel}      & \multicolumn{2}{c|}{non-parallel}  \\ \cline{3-6} 
                          &                                    & \multicolumn{1}{c}{Spec.}  & \multicolumn{1}{c||}{Dur.}   & \multicolumn{1}{c}{Spec.}  &   \multicolumn{1}{c|}{Dur.} \\ \hline\hline
\multirow{2}{*}{w2v2 {\sc BASE}}     & CSJ    & \multicolumn{1}{c}{1.19} & \multicolumn{1}{c||}{16.0} & \multicolumn{1}{c}{1.30} &  \multicolumn{1}{c|}{22.9}\\ 
                          & LibriSpeech                           & \multicolumn{1}{c}{1.16} & \multicolumn{1}{c||}{16.2} & \multicolumn{1}{c}{1.28} & \multicolumn{1}{c|}{23.3} \\ \hline
\multirow{1}{*}{w2v2 {\sc LARGE}}& LibriLight       & \multicolumn{1}{c}{\underline{\textbf{1.11}}} & \multicolumn{1}{c||}{15.9} & \multicolumn{1}{c}{\underline{\textbf{1.25}}} & \multicolumn{1}{c|}{22.2} \\ \cline{1-2} \cline{3-6} 
\multirow{2}{*}{HuBERT {\sc BASE}}   & CSJ                         & \multicolumn{1}{c}{1.15} &  \multicolumn{1}{c||}{\underline{\textbf{15.6}}}& \multicolumn{1}{c}{1.27} &  \multicolumn{1}{c|}{\underline{\textbf{22.0}}}\\ 
                          & LibriSpeech                          & \multicolumn{1}{c}{1.15} & \multicolumn{1}{c||}{16.0} & \multicolumn{1}{c}{1.29} &  \multicolumn{1}{c|}{23.0}\\ \hline 
 \multirow{1}{*}{HuBERT {\sc LARGE}}                         & LibriLight                          & \multicolumn{1}{c}{1.12} & \multicolumn{1}{c||}{16.1}& \multicolumn{1}{c}{1.27} & \multicolumn{1}{c|}{22.9}\\ \hline
\end{tabular}
 \vspace*{-1.mm} 
\end{table}
\vspace{-3.5mm}
\section{Conclusions}
\vspace{-2.5mm}
We proposed a zero-shot TTS method conditioned using an SSL model. Objective and subjective evaluations showed that the proposed method can generate utterances with higher reproduction than a conventional method using speaker recognition-based embedding, i.e., x-vector. The proposed method also enabled rhythm transfer by separate conditioning for each predictor. 

Although this paper focused on zero-shot TTS, we believe that the key idea of the proposed method would be easily applicable for other speech generation tasks, including fine-grained modeling for TTS~\cite{lee2019robust, sun2020fully} and VC considering speech rhythm~\cite{pmlr-v139-qian21b}. Applying the proposed embedding extractor to other speaker-related tasks not limited to speech generation, such as target speaker ASR~\cite{moriya22_interspeech} and target speech extraction~\cite{sato2022learning,sato21_interspeech}, is also for future work.



\newpage
\bibliographystyle{my_IEEEbib}
{\small \bibliography{strings,refs}}

\end{document}